\documentclass[%
aip,
preprint,
showpacs,
preprintnumbers,
amsmath,
amssymb,
]{revtex4-1}

\usepackage{graphicx}
\usepackage{dcolumn}
\usepackage{bm}
\usepackage{color}

\begin{document}

\title{Simultaneous polarization transformation and amplification of multi-petawatt laser pulses in magnetized plasmas }

\author{Xiaolong Zheng}
\affiliation{Key Laboratory for Laser Plasmas (MoE), School of Physics and Astronomy,
	Shanghai Jiao Tong University, Shanghai 200240, China}%
\affiliation{Collaborative Innovation Center of IFSA, Shanghai Jiao Tong University,
	Shanghai 200240, China}%

\author{Suming Weng}\email{wengsuming@gmail.com}%
\affiliation{Key Laboratory for Laser Plasmas (MoE), School of Physics and Astronomy,
	Shanghai Jiao Tong University, Shanghai 200240, China}%
\affiliation{Collaborative Innovation Center of IFSA, Shanghai Jiao Tong University,
	Shanghai 200240, China}%

\author{Zhe Zhang}
\affiliation{National Laboratory for Condensed Matter Physics, Institute of Physics, Chinese Academy of Sciences, Beijing 100190, China}%

\author{Hanghang Ma}
\affiliation{Key Laboratory for Laser Plasmas (MoE), School of Physics and Astronomy,
	Shanghai Jiao Tong University, Shanghai 200240, China}%
\affiliation{Collaborative Innovation Center of IFSA, Shanghai Jiao Tong University,
	Shanghai 200240, China}%

\author{Min Chen}
\affiliation{Key Laboratory for Laser Plasmas (MoE), School of Physics and Astronomy,
	Shanghai Jiao Tong University, Shanghai 200240, China}%
\affiliation{Collaborative Innovation Center of IFSA, Shanghai Jiao Tong University,
	Shanghai 200240, China}%

\author{Paul McKenna}
\affiliation{SUPA, Department of Physics, University of Strathclyde, Glasgow G4 0NG, UK}%

\author{Zhengming Sheng}\email{z.sheng@strath.ac.uk}%
\affiliation{Key Laboratory for Laser Plasmas (MoE), School of Physics and Astronomy,
	Shanghai Jiao Tong University, Shanghai 200240, China}%
\affiliation{Collaborative Innovation Center of IFSA, Shanghai Jiao Tong University,
	Shanghai 200240, China}%
\affiliation{SUPA, Department of Physics, University of Strathclyde, Glasgow G4 0NG, UK}%
\affiliation{Tsung-Dao Lee Institute, Shanghai 200240, China}%

\date{\today}

\begin{abstract}
With increasing laser peak power, the generation and manipulation of high-power laser pulses becomes a growing challenge for conventional solid-state optics due to their limited damage threshold.
As a result, plasma-based optical components which can sustain extremely high fields are attracting increasing interest. 
Here, we propose a type of plasma waveplate based on magneto-optical birefringence under a transverse magnetic field, which can work under extremely high laser power.
Importantly, this waveplate can simultaneously alter the polarization state and boost the peak laser power.
It is demonstrated numerically that an initially linearly polarized laser pulse with 5 petawatt peak power can be converted into a circularly polarized pulse with a peak power higher than 10 petawatts by such a waveplate with a centimeter-scale diameter.
The energy conversion efficiency of the polarization transformation is about $98\%$.
The necessary waveplate thickness is shown to scale inversely with plasma electron density $n_e$ and the square of magnetic field $B_0$, and it is about 1 cm for $n_e=3\times 10^{20}$ cm$^{-3}$ and $B_0=100$ T.
The proposed plasma waveplate and other plasma-based optical components can play a critical role for the effective utilization of multi-petawatt laser systems.
\end{abstract}

\maketitle

\section{Introduction}
Higher laser peak power has been continuously pursued since the laser was invented.
Thanks to development of laser technologies, especially chirped pulse amplification \cite{Strickland}, peak laser power at the petawatt (PW) level is now routinely achieved and a number of multi-PW lasers are built or under construction \cite{Danson}.
When such an ultra-high-power laser pulse is tightly focused, a laser peak intensity exceeding $10^{21}$ W/cm$^2$ can be achieved, which not only brings about many prospective applications but also becomes a unique tool to create extreme conditions for fundamental research \cite{MourouRMP,Gibbon}.
With increasing laser peak power, however, conventional solid-state optical components must be enlarged to avoid laser-induced damage \cite{Stuart}. In general, multi-PW laser systems require metre-scale optical components \cite{Cotel}, which are economically costly and technically challenging.

Plasma-based optical components offer an attractive solution to this issue, since all material will be at least partially ionized by an intense laser pulse and the resultant plasma can be used to manipulate the laser light.
The design and application of high-power lasers has the potential to be revolutionized by plasma-based optics \cite{Suckewer}, which may pave the way for the study of laser-matter interactions at unprecedented intensities.
So far, planar plasma mirrors are extensively used to the contrast of intense lasers \cite{Doumy,Thaury1}, while ellipsoidal plasma mirrors or flying plasma mirrors are used to focus or compress laser pulses toward extreme intensities \cite{Nakatsutsumi,Bulanov}.
Plasma-based optics are also extensively studied for other prospective applications, including plasma amplifiers \cite{Shvets,Andreev,Trines,Chiaramello,Turnbull2018}, plasma gratings or photonic crystals \cite{Sheng,WuPoP,WuAPL,YuJOSAB,Lehmann2016}, plasma optical modulators \cite{YuNC}, plasma apertures or shutters \cite{Gonzalez1,Gonzalez2,Wei}, plasma holograms \cite{Leblanc}, and cross beam energy transfer \cite{Michel,Moody}.
In particular, plasma gratings can act as polarizers or waveplates \cite{Michel2014,Turnbull2016,Lehmann2018} to control the polarization state of intense laser pulses.
Recently, magnetized plasmas have been proposed for the conversion of a linearly polarized (LP) laser pulse into two circularly polarized (CP) pulses by use of the Faraday effect under a strong longitudinal magnetic field \cite{Weng}.
The drawback of this scheme is that it reduces the peak power of the laser pulse by a factor of two.

In this work, we explore the birefringence effect of strong transverse magnetic fields induced in plasmas and propose to apply this effect to design plasma waveplates. In a plasma under a transverse magnetic field, the phase velocity of a light wave depends on the orientation of its electric field.
Therefore, the magnetized plasma can be used as a waveplate like a birefringent crystal. In particular, this magnetized plasma waveplate has two peculiar advantages in manipulating ultra-high-power laser pulses. Firstly, it is not susceptible to optical damage, as usually found for solid state optics.
Secondly, it can simultaneously alter the polarization state of an intense laser pulse and effectively boost the peak pulse power via  self-compression of the pulse.

\section{Theoretical analysis}
 
\begin{figure}
	\centering\includegraphics[width=0.8\textwidth]{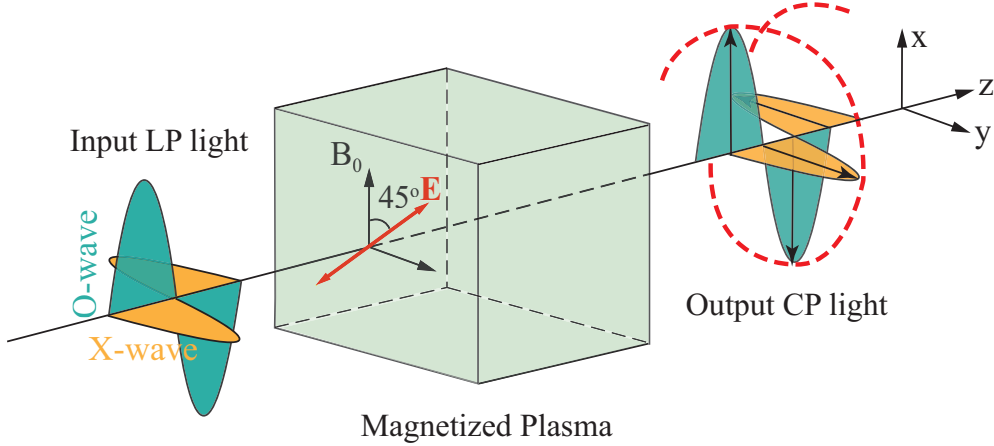}
	\caption{Sketch of the magnetized plasma quarter-wave plate. A LP light wave is incident into a plasma under a transverse magnetic field $\textbf{B}_0$. The incident wave can be equally divided into an ordinary sub-wave with $\textbf{E} \parallel \textbf{B}_0$ and an extraordinary sub-wave with $\textbf{E} \perp \textbf{B}_0$ if its polarization plane is oriented at $45^{\circ}$ with respect to $\textbf{B}_0$.
	A phase difference $\Delta \phi$ emerges between these two sub-waves because of their different phase velocities, and a CP wave can result when $\Delta \phi=\pi/2$.
	}\label{scheme}
\end{figure}

We first consider the propagation of a LP light wave in a plasma that is subjected to an external transverse magnetic field $\textbf{B}_0$. In general, this LP wave is a combination of two sub-waves: an ordinary wave with the electric field $\textbf{E} \parallel \textbf{B}_0$ and an extraordinary wave with $\textbf{E} \perp \textbf{B}_0$, hereinafter referred to as O-wave and X-wave, respectively.
The O- and X-waves have distinct dispersion relations as given respectively in the following \cite{FFChen}
\begin{eqnarray}
\frac{c^2 k^2}{\omega^2} &=& \frac{c^2}{v_{\phi,O}^2} = 1 - \frac{\omega_p^2}{\omega^2},  \\
\frac{c^2 k^2}{\omega^2} &=& \frac{c^2}{v_{\phi,X}^2} = 1 - \frac{\omega_p^2}{\omega^2} \frac{\omega^2-\omega_p^2}{\omega^2-\omega_p^2-\omega_c^2},
\end{eqnarray}
where $\omega$ and $k$ are the wave angular frequency and wavenumber, $\omega_p=(n_e e^2/\epsilon_0 m_e)^{1/2}$ and $\omega_c=eB_0/m_e$ are respectively the plasma frequency and the electron cyclotron frequency, and $n_e$ is the electron number density. The above equations define the distinct phase velocities $v_{\phi,O}$ and $v_{\phi,X}$ for the O- and X-waves, respectively.
Subsequently, a phase difference $\Delta \phi$ will be induced between the O- and X-waves, which makes the incident LP wave elliptically polarized.
Therefore, the magnetized plasma can be considered as a wave retarder or waveplate.
A notable case is that an incident wave LP at $45^{\circ}$ with respect to $\textbf{B}_0$ can be converted into a CP wave as shown in Fig. \ref{scheme} when the phase difference
\begin{equation}
\Delta \phi= 2\pi \frac{ d}{\lambda} (\frac{c}{v_{\phi,O}}-\frac{c}{v_{\phi,X}})=\frac{\pi}{2}, \label{PhaseD}
\end{equation}
where $d$ is the propagation distance in the plasma and $\lambda$ is the wavelength (see \textcolor{blue}{Visualization 1} for the time evolution of the electric field vector in this polarization transformation). In the case $\omega_p \ll \omega$ and $\omega_c \ll \omega$, the phase difference can be simplified to $\Delta \phi \simeq \pi (d/\lambda) (n_e/n_c)  (e B_0/m_e \omega)^2$, where $n_c= \epsilon_0 m_e \omega^2/e^2$  is the critical plasma density. 

This magneto-birefringence effect in plasmas is analogous to the Voigt effect in gases found in 1902 or the Cotton-Mouton effect in liquids discovered in 1907 \cite{Voigt}. 
{However, this effect is usually applied only for the polarimetric diagnosis of magnetic fields \cite{Segre}. To the best of our knowledge, it is the first time to propose a concept of magnetized plasma waveplate based on this effect for high power laser pulses.}
Above all, magnetized plasmas as a restoring medium have unique strengths in manipulating intense laser pulses. Importantly, an intense laser pulse can be further amplified by nonlinear effects in a plasma \cite{Shvets,Andreev,Trines,Chiaramello,Turnbull2018}.
In particular, a weakly relativistic laser pulse can be intensified via the process of self-compression in a plasma \cite{Shorokhov}.
In a plasma, the envelope of an intense laser pulse can be approximately described by the nonlinear Schr\"{o}dinger equation (NLSE) \cite{Ren}
\begin{equation}
2i \frac{\omega^2}{\omega_p^2} \frac{v_g^3}{c^3} \frac{\partial a}{\partial \tau} + \frac{\omega^2}{\omega_p^2} \frac{v_g^2}{c^2}\nabla_\bot^2a+
\frac{\partial^2 a}{\partial \psi^2} + \frac{v_g^2}{c^2}(1-\frac{1}{\gamma})=0, \label{NLSE3D}
\end{equation}
where $\tau=z\omega/c$, $\psi=(t-z/v_g)\omega$, $v_g=kc^2/\omega$, the normalized electromagnetic vector potential $\textbf{a}=e\textbf{A}/m_ec^2$ , {and the electron Lorentz factor $\gamma \simeq \sqrt{1+a^2/2}$ in a LP laser pulse}. Here the effect of the external magnetic field can be neglected as long as $\omega_c \ll \omega$. The transverse self-focusing becomes obvious at a distance $z=Z_R(P/P_c-1)^{-1/2}$ \cite{Gibbon}, where $Z_R=\pi r_0^2/\lambda$ is the Rayleigh length, $r_0$ is the pulse waist, $P$ is the pulse power, and $P_c\simeq17.5n_c/n_e$ GW is the critical power for relativistic self-focusing. Since $z\propto r_0^2$, the self-focusing is usually negligible within the Rayleigh length when the laser pulse is not tightly focused. Consequently, Eq. (\ref{NLSE3D}) is descended to the one-dimensional (1D) NLSE \cite{Shorokhov,Ren}
\begin{equation}
2i \frac{\omega^2}{\omega_p^2} \frac{v_g^3}{c^3} \frac{\partial a}{\partial \tau} + \frac{\partial^2 a}{\partial \psi^2} + \frac{v_g^2}{c^2}(1-\frac{1}{\gamma})=0, \label{NLSE}
\end{equation}
which describes the pure longitudinal self-compression.

\section{Simulation verification}

To test the performance of magnetized plasma waveplates, we perform a series of particle-in-cell (PIC) simulations using the Osiris code \cite{Fonseca}.
In each simulation, a laser pulse with a wavelength $\lambda=1$ $\mu\texttt{m}$ is incident along the $z$-axis into a magnetized plasma that is located at $0 \leq z \leq d_0$.
The moving window is employed with a simulation box moving along the $z$ axis at the speed of light in vacuum. In 1D simulations, the spatial and temporal resolutions are $\Delta z = \lambda/40$ and $\Delta t \simeq \Delta z / c$, and each cell has 8 macroparticles. In the 3D simulation, the spatial resolutions are $\Delta z = \lambda/16$ and $\Delta x = \Delta y = 50 \lambda$, the temporal resolution is $\Delta t \simeq \Delta z /c$, and each cell has four macroparticles.
The initial pulse is assumed to be LP at $+45^{\circ}$ with respect to the $x$-axis, and the external magnetic field $\textbf{B}_0$ is along the $x$-axis. The polarization state of the laser pulse is described by the Stokes parameters \cite{Saleh}: $I=|E_x|^2+|E_y|^2$, $Q=|E_x|^2-|E_y|^2$, $U=2 \texttt{Re} {\{E_x^*E_y}\}$ and $V= 2 \texttt{Im} {\{E_x^*E_y}\}$, where $E_x$ ($E_y$) is the complex amplitude of the electric field along the $x$ ($y$) axis, and $E_x^*$ is the complex conjugate of $E_x$. The parameter $I$ denotes the intensity regardless of polarization, $Q$ the LP along the $x$ ($+$) or $y$ ($-$) axes, $U$ the LP at $+45^{\circ}$ ($+$) or $-45^{\circ}$ ($-$) from the $x$ axis, $V$ the right-handed ($+$) or left-handed ($-$) CP. For comparison, all parameters are normalized to the initial peak intensity $I_0$.

\begin{figure}
    \centering\includegraphics[width=0.8\textwidth]{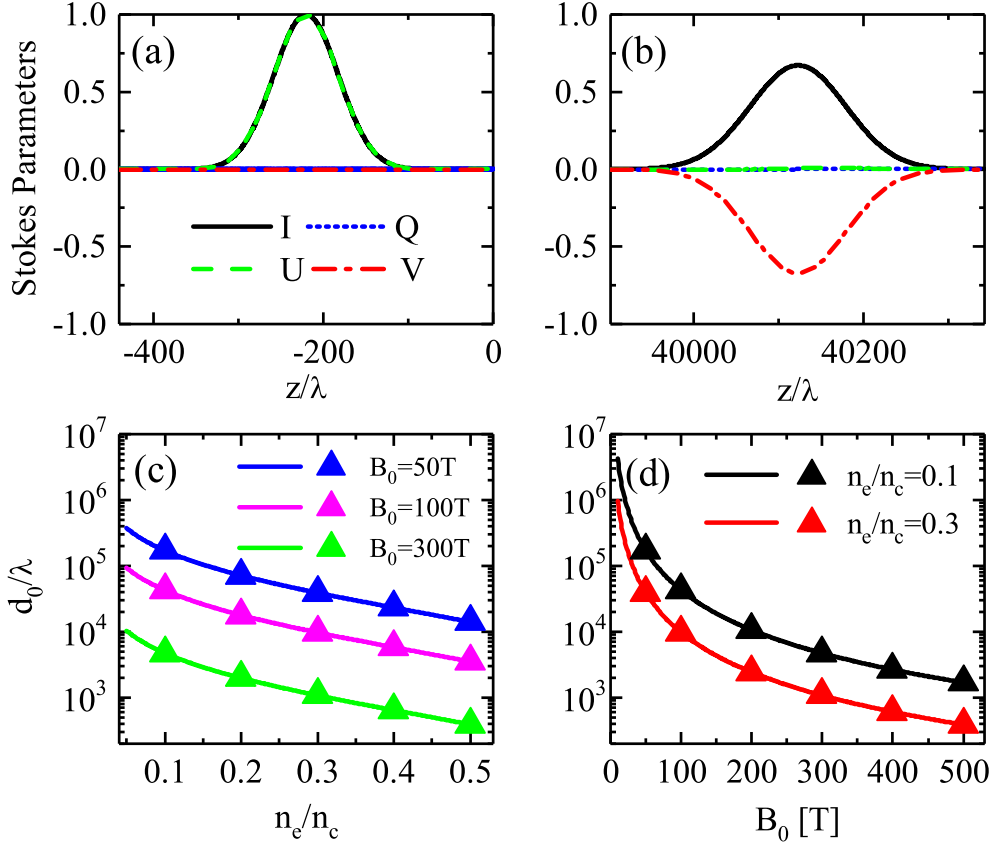}
	\caption{Stokes parameters of a laser pulse (a) before and (b) after a magnetized plasma quarter-wave plate with $d_0\simeq 3.9$ cm, $n_e/n_c=0.3$, and $B_0=50$ T.
	To be a quarter-wave plate, the required plasma thickness $d_0$ as a function of (c) plasma density $n_e$ and (d) magnetic field $B_0$, where the PIC simulation results (triangles) are in good agreement with the predictions by Eq. (\ref{PhaseD}) (solid curves).
	In the simulation, the laser pulse initially has a FWHM duration of 300 fs and a relatively weak intensity ($a_0=0.001$).
	}\label{figNonrel}
\end{figure}

Figure \ref{figNonrel} displays 1D PIC-simulation results for a weak laser pulse with an amplitude $a_0=0.001$ ($I_0 \simeq 10^{12}$ W/cm$^2$).
Figures \ref{figNonrel} (a) and (b) show the Stokes parameters of the pulse before and after it passes through a magnetized plasma with $d_0 \simeq 3.9$ cm, $n_e/n_c=0.3$ and $B_0 \simeq 50$ T ($eB_0/m_e\omega=0.005$). According to Eq. (\ref{PhaseD}), this magnetized plasma will induce a phase difference $\Delta \phi \simeq \pi/2$ and thus act as a quarter-wave plate. Therefore, the laser polarization is converted from LP at $+45^{\circ}$ $(U/I\simeq +1)$ to left-handed CP $(V/I\simeq -1)$.
Further, Figs. \ref{figNonrel} (c) and (d) show that the required plasma thickness $d_0$ scales inversely with the plasma density $n_e$ and the square of the magnetic field $B_0$, as predicted by Eq. (\ref{PhaseD}).
In the simulations with weak laser pulses, the degrees of circular polarization $|V|_{\max}/I_{\max}$ of the output pulses always exceed $99\%$.
However, the output peak intensities are, to some extent, smaller than the input values due to the dispersive broadening. For example, it is about $0.67I_0$ in the case shown in Fig. \ref{figNonrel}(b). 

\begin{figure}
	\centering\includegraphics[width=0.8\textwidth]{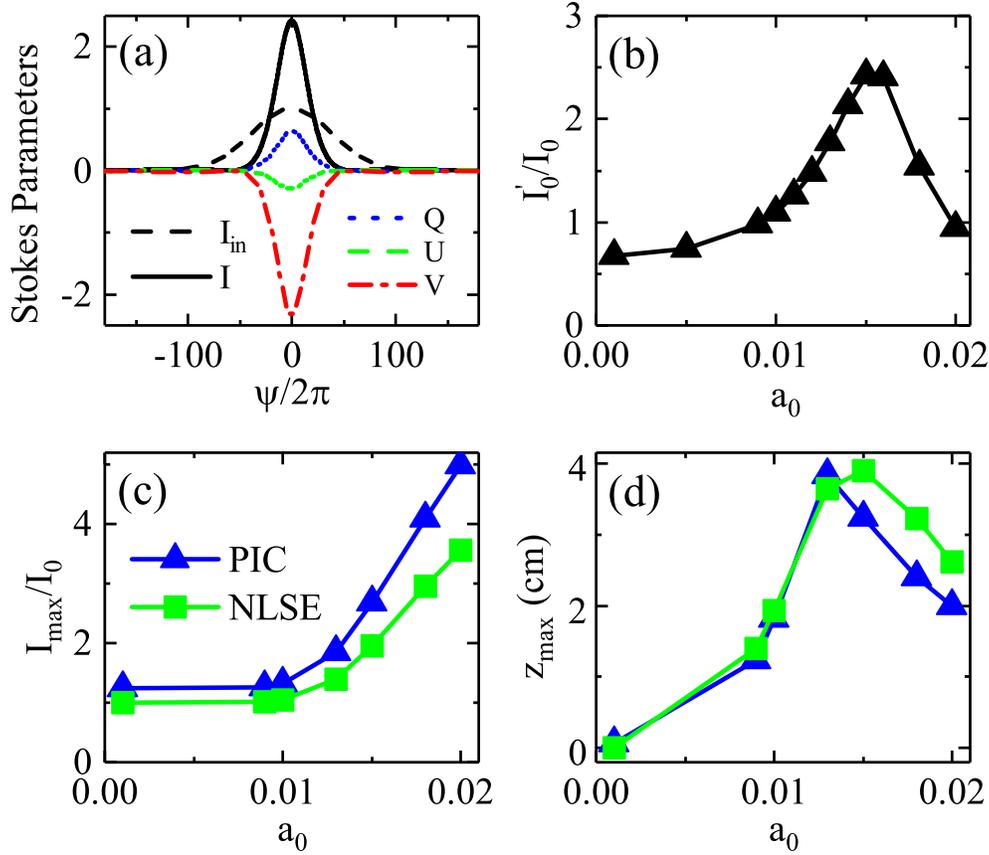}
	\caption {(a) Stokes parameters for a weakly relativistic laser pulse with $a_0=0.015$ after it passes through the magnetized plasma, where the input intensity profile $I_\texttt{in}$ is also drawn for comparison;
	(b) the output peak intensity $I_0'$ as a function of the initial amplitude $a_0$.
	(c) The maximum peak intensity $I_{\max}$ can be achieved during the laser propagation in the plasma and (d) the propagation distance $z_{\max}$ where this maximum is achieved as functions of $a_0$; where the PIC simulation results (triangles) are in good overall agreement with the predictions by NLSE Eq. (\ref{NLSE}) (squares).
	Except $a_0$, all other parameters are the same as those in Fig. \ref{figNonrel} (b).
	}\label{figRel}
\end{figure}

When increasing the laser pulse intensity such that it is weakly relativistic, we find that the dispersive broadening can be easily compensated by the pulse self-compression.
As illustrated in Fig. \ref{figRel}(a), a laser pulse with $a_0=0.015$ is intensified by a factor of $2.4$ in another 1D PIC-simulation using the same magnetized plasma as used for the results presented in Fig. \ref{figNonrel}(b). Meanwhile, the output pulse maintains a high degree of circular polarization ($\sim 95\%$).
Fig. \ref{figRel}(b) shows that the output peak intensity increases with the increasing $a_0$ when $a_0\leq 0.015$. This is because the self-compression becomes more and more significant with increasing laser intensity \cite{Shorokhov}.
Figure \ref{figRel}(c) shows that the maximum achievable amplification factor $I_{\max}/I_0$ of the pulse increases with increasing $a_0$.
For cases $a_0 > 0.015$, however, the self-compression becomes so fast that the maximum amplification factor will be achieved for a decreasing propagation distance $z_{\max}$ with the increasing $a_0$ as shown in Fig. \ref{figRel}(d).
Therefore, the maximum amplification of the pulse will be achieved in the interior of the plasma.
Consequently, in the case $a_0 > 0.015$ the final output peak intensity from the magnetized plasma in Fig. \ref{figRel}(b) will be lower than the maximum achievable value shown in Fig. \ref{figRel}(c).
The small difference between the maximum achievable peak intensities from the numerical analysis and PIC simulations in Fig. \ref{figRel}(c) may be due to the simplifications in the deduction of Eq. (\ref{NLSE})\cite{Ren}. Further, it is worth mentioning that  the magnetic field used in this work has no obvious influence on the pulse compression since $\omega_c \ll \omega_p$ \cite{Wilson2017,Wilson2019}. Therefore, we ignore the effect of magnetic field in the theoretical model for the pulse self-compression, i.e., Eq. (\ref{NLSE}).

\begin{figure}
	\centering\includegraphics[width=0.8\textwidth]{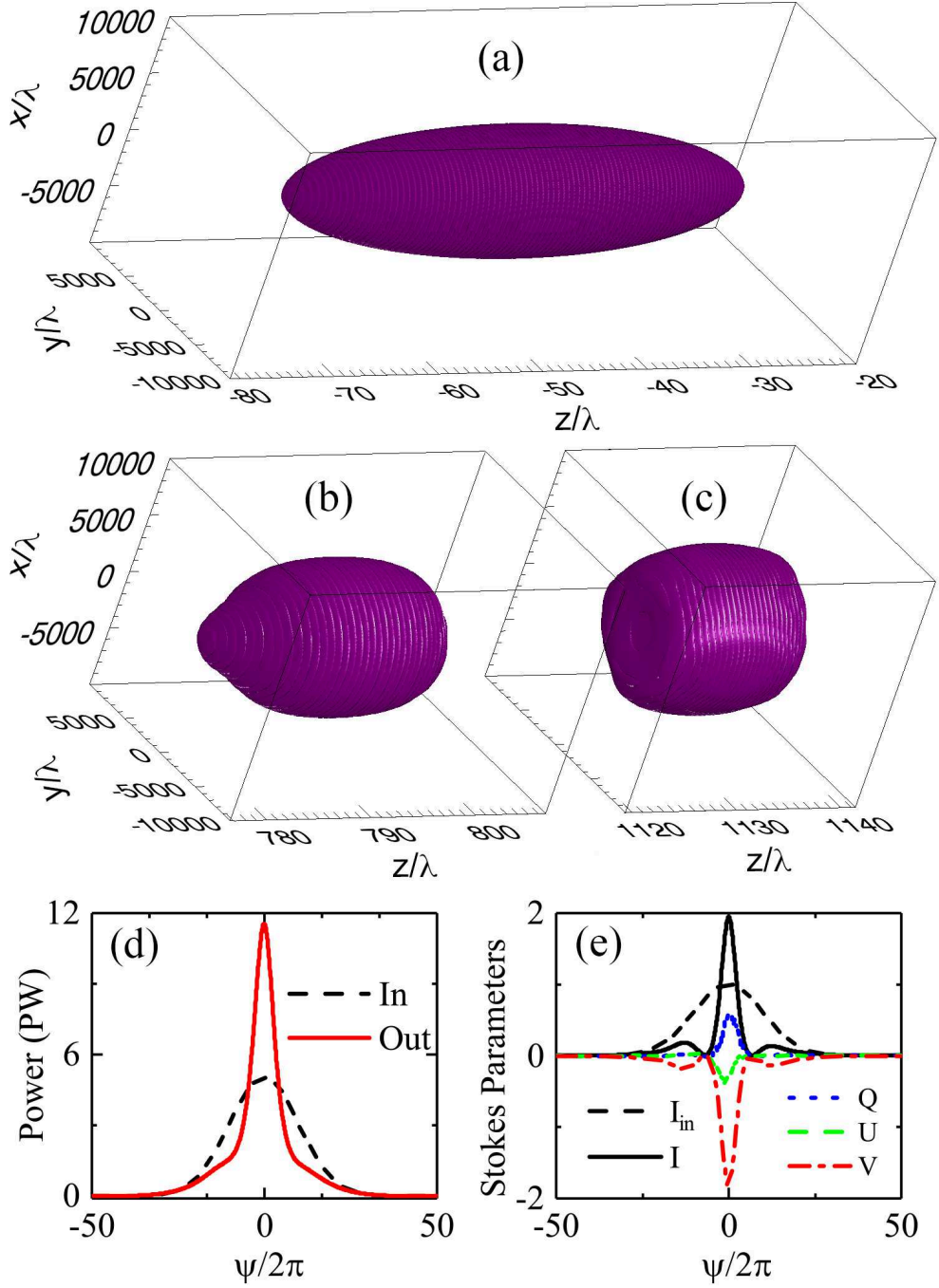}
	\caption {Isosurfaces of the electromagnetic field energy density $E^2+B^2=\mathcal{E}_{0}/10$ (a) before ($t=-50$), (b) during ($t=950$), and (c) after ($t=1350$) the laser pulse interaction with a magnetized plasma, where $\mathcal{E}_{0}$ is the maximum of the initial electromagnetic energy density;
		(d) the pulse power as a function of the variable $\psi$ for the input and output pulse, respectively;
		(e) Stokes parameters on the $z$-axis of the laser pulse after it passes through the magnetized plasma, where the input intensity profile $I_\texttt{in}$ is also drawn for comparison.
	}\label{fig3D}
\end{figure}

{The above results indicate that the higher is the laser intensity, the faster the self-compression occurs. 
So it seems that the magnetized plasma quarter waveplate should have an extremely thin thickness for laser pulses at relatively high intensities. Correspondingly, the required magnetic field should be extremely strong according to Eq. (\ref{PhaseD}). 
However, it is important to note that an intense laser pulse can experience a periodic compression and decompression, in which the peaks of the pulse amplification appear periodically \cite{Shorokhov}. 
Therefore, the plasma waveplate thickness can be set as the propagation distance where the second or subsequent amplification peak is achieved. 
Correspondingly, the magnetic field strength of plasma quarter waveplate for intense laser pulses might be greatly reduced to a more accessible level.
}

The capability of the magnetized plasma waveplate for laser pulse intensification has also been demonstrated via a 3D PIC simulation as shown in Fig. \ref{fig3D}, where a shorter laser pulse duration 80 fs and a stronger magnetic field $B_0=300$ T are invoked to reduce the computation cost.
For $B_0=300$ T, the required thickness of the plasma with $n_e/n_c=0.3$ will be greatly reduced to $d_0\simeq 1083\lambda$ as predicted by Eq. (\ref{PhaseD}).
Accordingly, the initial laser amplitude increases up to $a_0=0.1$ in order to speed up the pulse compression and achieve an obvious intensification at $d=d_0$.
The strong self-compression of the laser pulse is clearly visible in Figs. \ref{fig3D}(a), (b) and (c) (see \textcolor{blue}{Visualization 2} for the entire self-compression process), where a large pulse waist $r_0=4800\lambda$ is set to gain an initial peak power $P\simeq5$ PW.
The pulse compression results in an increasing power. As shown in Fig. \ref{fig3D}(d), the peak power is boosted from 5 PW up to 11.5 PW by the magnetized plasma waveplate.
Correspondingly, the peak intensity is enhanced by about a factor of 2 as shown in Fig. \ref{fig3D}(e), which also illustrates that the output pulse has a high degree of circular polarization ($\sim 94\%$).
The energy conversion efficiency of the polarization transformation is as high as $98\%$.
The resultant ultra-high-power CP pulse, focused properly, could be extensively applied to laser-driven ion acceleration and ultra-bright X-ray radiation \cite{Scullion,Yu,Bin,Shen}.

\section{Discussion}

Although circularly polarized laser pulses with peak powers as high as 10 petawatts have already been assumed broadly in recent theoretical studies \cite{Scullion,Yu,Bin,Shen}, the generation of such pulses is still one of the outstanding problems in the field of high-power laser science.
Nowadays, the generation of circularly-polarized laser pulses still relies on the use of conventional crystal quarter-wave plates. For a 10 petawatt laser pulse, however, the diameter of the crystal waveplate must be of the order of metre-scale to avoid laser-induced damage. It is presently not possible to manufacture such a large-diameter crystal waveplate of the required thickness and optical quality.
The magnetized plasma waveplate introduced above provides an alternative approach to manipulate the polarization state of a laser pulse. In the encounter with an ultra-high-power laser pulse, such a magnetized plasma waveplate has two advantages over the conventional waveplate made from a birefringent crystal.
Firstly, the magnetized plasma waveplate can be more compact since it can sustain an intense laser pulse.
As illustrated in the above 3D simulation, a plasma waveplate on the centimeter scale (a waist of 4.8 mm) enables the transmission of a 10 PW laser pulse. 
Secondly, the magnetized plasma waveplate can compress an intense laser pulse and boost its peak power. This is of great benefit to ultra-high-power laser systems since it may partially liberate the role of the grating compressor in directly achieving the 10 PW level laser pulses.

\begin{figure}[h]
	\centering\includegraphics[width=0.8\textwidth]{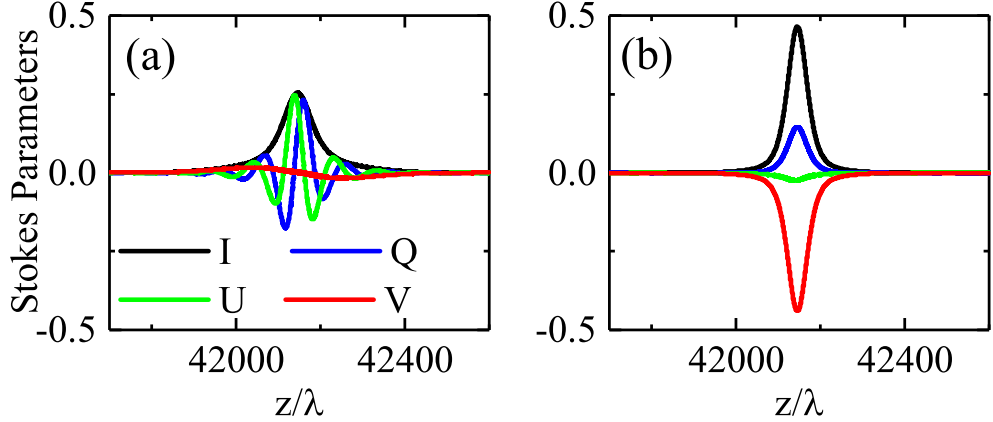}
	\caption {
	Stokes parameters of a laser pulse after it passes through a plasma under (a) a longitudinal magnetic field or (b) a transverse magnetic field. Except for the magnetic field direction, all other parameters used in (a) and (b) are the same. The magnetic field strength is $B_0 = 50$ T. The initial laser pulse is LP with  $a_0 = 0.025$ and a duration of 100 fs. The plasma has a uniform density $n_e= 0.3n_c $, and a thickness $d \simeq 3.9$ cm. 
	With $d \simeq 3.9$ cm, the magnetized plasma in (b) plays a role as a quarter-wave plate according to Eq. (\ref{PhaseD}). Therefore, the output laser pulse is CP with $ |V/I| \simeq 1$ as shown in (b).
	With  $d \simeq 3.9$ cm, the magnetized plasma in (a) should be long enough to separate the left-handed and right-handed CP sub-pulses according to Eq. (4) in Ref. \cite{Weng}. However, no pulse splitting is observed in (a) since $B_0 = 50$ T is below the threshold for the pulse splitting. 
	}\label{fig5}
\end{figure}

As magnetized plasmas have been proposed for the polarization transformation of high-power laser pulses previously\cite{Weng}, we would like to emphasize that the polarization transformation scheme proposed in the current paper is based on completely new underlying physics and has a few prominent advantages over the schemes already proposed.
The scheme proposed here utilizes the birefringence in a plasma under a transverse magnetic field, while the scheme in Ref. \cite{Weng} is based on the chirality (i.e. circular dichroism) in a plasma under a longitudinal magnetic field.
The new underlying physics of the current scheme gives it two advantages over the scheme in Ref. \cite{Weng}.
Firstly, an initial LP laser pulse will be split into a left-handed CP and a right-handed CP laser pulses by using the scheme in Ref. \cite{Weng}. Therefore, the peak power of the laser pulse will be reduced by a factor of two in general.
In contrast, neither the pulse splitting nor the significant reduction in the peak power occurs in the current scheme since it transforms an initial LP laser pulse into another single CP pulse directly. The slight reduction in the peak power due to the dispersive broadening could be compensated by the relativistic self-compression of the pulse. 
Secondly, the electron cyclotron frequency ($\omega_c=eB/m_e$) in the magnetic field should be larger than the frequency spread ($\Delta \omega$) of the laser pulse in order to guarantee that the pulse splitting is faster than the dispersive broadening by using the scheme in Ref. \cite{Weng}. 
That is to say, the magnetic field is required to be stronger than a threshold for the pulse splitting
\begin{equation}
B \ge  B_{\min} =\frac{ m_e \Delta \omega }{e}.  \label{minB}
\end{equation}
This usually demands an ultra-strong magnetic field for an ultrashort laser pulse. In contrast, there is no magnetic field strength constraint in the current scheme since it demands no pulse splitting in principle.
As illustrated in Fig. \ref{fig5}(a), using the scheme in Ref. \cite{Weng}, an initial 100 fs LP laser pulse cannot be split into two CP laser pulse no matter how far it propagates in a plasma under a longitudinal magnetic field $B=50$ T. Since the frequency spread $\Delta \omega \simeq 0.015 \omega$ is larger than the electron cyclotron frequency $\omega_c=eB/m \simeq  0.005\omega$. In this case, the dispersive broadening dominates over the pulse splitting\cite{Weng}.
In contrast, Fig .\ref{fig5}(b) demonstrates that this initial LP laser pulse can be transformed into a  CP laser pulse by using the current scheme that is based on the birefringence under a transverse magnetic field with the same strength.

\begin{figure}[h]
	\centering\includegraphics[width=0.4\textwidth]{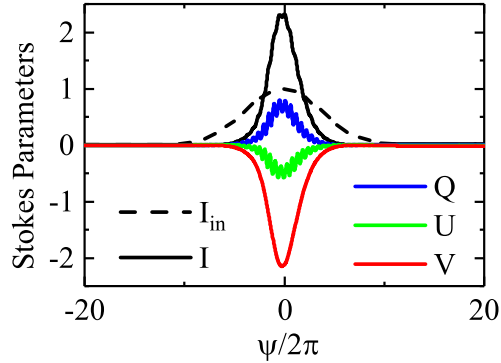}
	\caption {Stokes Parameters of a short laser pulse after it passes through a magnetized plasma with $d_0=0.474$ cm, $n_e=0.1n_c$ and $B=300$ T, where the input intensity profile $I_\texttt{in}$ is also drawn for comparison. The initial LP laser pulse has $a_0=0.14$ and $t_p=30$ fs.}\label{fig6}
\end{figure}

{For an ultrashort laser pulse with a broad spectrum, however, magnetized plasma quarter-wave plates also suffer from a similar problem of conventional solid-state quarter-wave plates. That is, the resultant phase difference $\Delta \phi$ between O-wave and X-wave is exactly equal to $\pi/2$ only for the central wavelength. 
Therefore, it is difficult to convert  an initial ultrashort LP laser pulse into a CP pulse with a high level of circular polarization. 
However, magnetized plasma quarter waveplates may partially alleviate this problem because it can compress a laser pulse as well as alter its polarization state. 
Therefore, an ultrashort CP laser pulse can be obtained from an initially relatively long LP laser pulse. Although the final pulse duration is ultrashort in this case, the spectrum that determined by the initial duration is not so broad. So the level of circular polarization for the final CP pulse could be relatively high.
As illustrated in Fig. \ref{fig6}, an initial LP laser pulse with $t_p=30$ fs can be converted to a CP laser pulse whose FWHM duration is only about 10 fs and circular polarization level $|V/I|$ is as high as 92\%..}

It is worth pointing out that this proposed magnetized plasma waveplate also has some limitations on the laser and plasma parameters, although plasmas are usually considered to be free from laser-induced damage. 
As suggested previously \cite{Weng}, it is important to set the laser intensity $a_0 \sim 0.1$ and the plasma density $n_e \ll n_c$. 
The conditions $a_0\sim 0.1$ and $n_e \ll n_c$ guarantee the collisionless damping small. We find that the energy loss due to the collisionless damping is maintained at a level $\sim 2\%$ as long as $a_0 \le 0.1$ and $n_e \le  0.3n_c$ in the simulations.
{Taking into account the collisional damping, there will be some additional limitations on the plasma density and/or magnetic field strength.
The losses due to the inverse bremsstrahlung,  which are not included in our PIC simulations, can be estimated as $K_{ib}=1-\exp{(-\kappa_{ib} d_0)}$ \cite{EliezerBook}, where $d_0$ is the thickness of magnetized plasma waveplate and $\kappa_{ib} \simeq  \nu_{ei} (n_e/n_c)^2(1-n_e/n_c)^{-1/2}/c$ is the spatial damping rate. 
For $a_0 \sim 0.1$, the effective electron-ion collision frequency is given by $\nu_{ei}\simeq Z_i e^4 n_e \ln \Lambda / (4 \pi \epsilon_0^2 m_e^2 v_{eff}^3)$ \cite{Gibbon, WengPRE}, with the effective electron velocity $v_{eff} = (v_{te}^2 + v_{os}^2)^{1/2} \simeq a_0 c$, the ionization state $Z_i$ , the Coulomb logarithm $\ln \Lambda$,  the electron thermal velocity $v_{te}$ and oscillatory velocity in the laser field $v_{os} \simeq a_0c$. 
Therefore, we can estimate that the thickness of magnetized plasma waveplate should be thinner than 0.3 cm in order to maintain the collisional losses $<10\%$ in the case $a_0=0.1$ and $n_e=0.3n_c$. 
According to Eq. (\ref{PhaseD}), this demands a magnetic field stronger than 200 T, which might be generated only by laser-driven capacitor coils \cite{Fujioka,Law,Tikhonchuk}. 
Fortunately, the collisional damping rate $\kappa_{ib}$ decreases quickly with a decreasing plasma density. 
Consequently, the limitation upon the magnetic field strength can be greatly relaxed with a lower plasma density. 
For instance, we find that a magnetic field $>70$ T (correspondingly $d_0<9$ cm) is enough to maintain the collisional losses $<10\%$ in the case $a_0=0.1$ and $n_e=0.1n_c$.}

{For a high power laser pulse, its waist will be very large if it is not focused tightly.
For instance, the waist  $r_0 \simeq 4800 \lambda$ in the case $a_0=0.1$ for a 5 PW laser pulse. Correspondingly, the Rayleigh length $z_R=\pi r_0^2/\lambda \simeq 7.24 \times 10^7 \lambda$. This extremely long Rayleigh length guarantees that multi-dimensional effects such as  relativistic self-focusing and  transverse filamentation  instabilities do not occur strongly before the polarization transformation of the laser pulse is completed. We have performed a few 2D test simulations with $r_0 \simeq 4800 \lambda$, and no obvious transverse instabilities is observed within a propagation distance up to a few centimeters. 
There is also no obvious Raman scattering instability in the performed 1D and 2D test simulations with $n_e<n_c/4$. This may be because the employed laser pulses are relatively short (on the order of 100 fs) and weakly relativistic ($a_0 \le 0.1$).}


It is worth mentioning that it is extremely high demand for computational resources to carry out 3D PIC simulations with high transverse resolutions $\delta x, \delta y < \lambda$ due to the huge focal spot size of the laser pulse ($r_0=4800\lambda$).
To make sure there is no problem to use such large resolutions in the transverse direction, we have also done a few 2D test simulations with the transverse resolution ranging from $\lambda/8$ to $50 \lambda$, and we find no obvious difference between the simulation results as long as the transverse resolution is much smaller than the laser pulse waist $r_0$.

\section{Conclusion}

In conclusion, we have shown that a plasma slab under a transverse magnetic field can act as a waveplate due to magnetic-field induced birefringence.
This magnetized plasma waveplate can not only sustain intensities several orders of magnitude higher than a crystal waveplate but also boost the peak power of the laser pulse.
Therefore, this scheme could be developed as a powerful means to manipulate the polarization state of ultra-high-power laser pulses.
Such a magnetized plasma waveplate should make it feasible to achieve 10 PW circularly polarized laser pulses that are technologically challenging by other means.
The magnetized plasma waveplate can be considered as a new addition to the plasma-base class of optical components, which have the potential to revolutionize the design and application of ultra-high-power lasers.

\section*{Funding}
National Natural Science Foundation of China (Grant Nos. 11675108, 11655002, 11721091, 11535001, and 11405108), National Basic Research Program of China (Grant No. 2013CBA01504), National 1000 Youth Talent Project of China, Science Challenge Project (No.TZ2018005) and EPSRC (Grant No. EP/R006202/1).

\section*{Acknowledgments}
Simulations have been carried out on the Pi supercomputer at Shanghai Jiao Tong University. Data
associated with research published in this paper can be accessed at https://doi.org/10.15129/8d4b5e09-
5680-4b3d-8501-0cba442892a3.

\end{document}